%% file: main.tex
\documentclass[sigconf,10pt]{acmart}
\usepackage{enumitem}

\usepackage[dvipsnames]{xcolor}
\usepackage{soul}
\usepackage{minted}
\usepackage{hyperref}

\input{macro}

\AtBeginDocument{%
  }

\setcopyright{acmlicensed}
\acmYear{2025}\copyrightyear{2025}
\setcopyright{cc}
\setcctype[4.0]{by}
\acmConference[WiNTECH '25]{ACM Workshop on Wireless Network Testbeds, Experimental evaluation \& Characterization}{November 4--8, 2025}{Hong Kong, China}
\acmBooktitle{ACM Workshop on Wireless Network Testbeds, Experimental evaluation \& Characterization (WiNTECH '25), November 4--8, 2025, Hong Kong, China}
\acmDOI{10.1145/3737895.3768303}
\acmISBN{979-8-4007-1972-1/25/11}

\title{{\namebf}: Bridging LLMs and IoT Systems Through Model Context Protocol}

\author{Ningyuan Yang, Guanliang Lyu, Mingchen Ma, Yiyi Lu, Yiming Li, Zhihui Gao, Hancheng Ye, Jianyi Zhang, Tingjun Chen, Yiran Chen}
\affiliation{%
  \institution{\vspace{0.5ex} Department of Electrical and Computer Engineering, Duke University\vspace{0.5ex}}
  \country{}
}

\renewcommand{\shortauthors}{Yang et al.}

\begin{document}

\begin{abstract}

The integration of Large Language Models (LLMs) with Internet-of-Things (IoT) systems faces significant challenges in hardware heterogeneity and control complexity. The Model Context Protocol (MCP) emerges as a critical enabler, providing standardized communication between LLMs and physical devices. We propose IoT-MCP, a novel framework that implements MCP through edge-deployed servers to bridge LLMs and IoT ecosystems. To support rigorous evaluation, we introduce \textit{IoT-MCP Bench}, the first benchmark containing 114 Basic Tasks (e.g., ``What is the current temperature?'') and 1,140 Complex Tasks (e.g., ``I feel so hot, do you have
any ideas?'') for IoT-enabled LLMs. Experimental validation across 22 sensor types and 6 microcontroller units demonstrates IoT-MCP's 100\% task success rate to generate tool calls that fully meet expectations and obtain completely accurate results, 205ms average response time, and 74KB peak memory footprint. This work delivers both an open-source integration framework (\url{https://github.com/Duke-CEI-Center/IoT-MCP-Servers}) and a standardized evaluation methodology for LLM-IoT systems.
\end{abstract}

\ccsdesc[500]{Hardware~Sensors and actuators}

\keywords{Internet-of-Things (IoT), Model Context Protocol (MCP), Generative AI}


\maketitle

\section{Introduction}

The proliferation of Internet-of-Things (IoT) systems has created unprecedented opportunities for physical world digitization, yet managing heterogeneous devices remains challenging~\cite{atzori2010internet, chen2018maximizing}. 
Large Language Models (LLMs)~\cite{vaswani2017attention, touvron2023llama} offer transformative potential for IoT systems by enabling natural language interaction with IoT devices, but also require intelligent and robust interfaces to bridge the digital and physical worlds. 
The recently emerged Model Context Protocol (MCP)~\cite{hou2025modelcontextprotocolmcp} addresses this need through its protocol-agnostic design and extensible tool framework, empowering LLMs to perceive, influence, and interact with IoT systems.
With industry adoption accelerating~\cite{IntroducingModelContext, ModelContextProtocol}, rigorous MCP evaluation in the context of different IoT systems becomes critical.
However, real-world IoT deployments face significant challenges: extreme hardware heterogeneity complicates scalability; low-latency requirements demand careful data/connection balancing; and performance assessment lacks standardized metrics, unlike structured domains.

\begin{figure}[!t]
    \centering
    \includegraphics[width=1\linewidth]{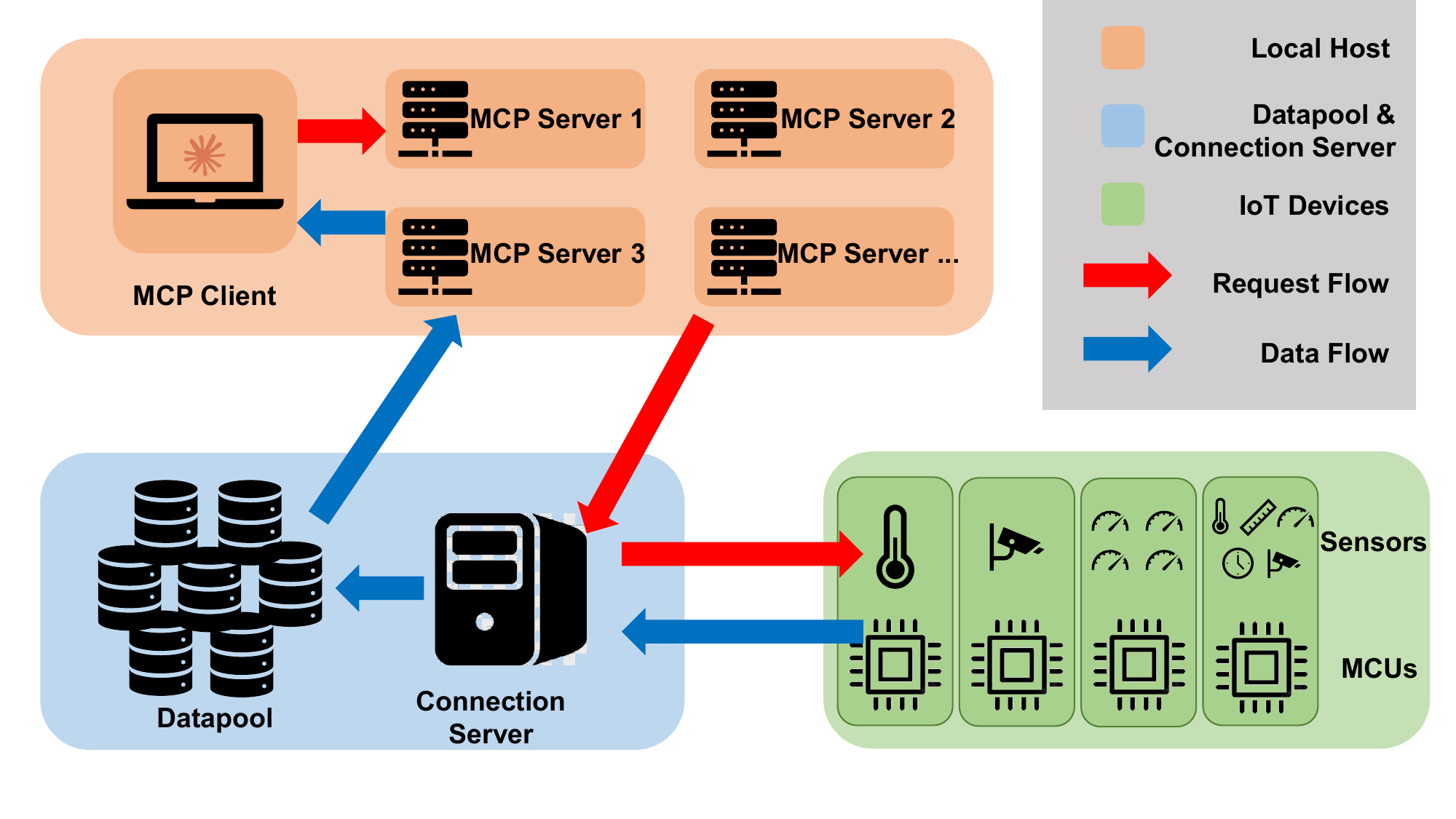}
    \caption{Workflow of the developed IoT-MCP framework, illustrating the request and data flow between the MCP servers and clients on the Local Host (Orange), Datapool and Connection Server (Blue), and IoT Devices (Green).}
    \label{fig:iotmcp-overview}
\end{figure}

To address these challenges, in this paper, we introduce \textbf{IoT-MCP} (see Figure~\ref{fig:iotmcp-overview}), a comprehensive framework that enables LLM to interact with IoT systems through standardized MCP interfaces.
{\name} adopts a decoupled architecture design and can be divided into three modules. \emph{Local Host} implements a lightweight MCP architecture to ensure stable and efficient interaction between users and LLM. \emph{Datapool and Connection Server} centrally manages data request interactions between all MCP servers and MCUs, establishing efficient and long connections. \emph{IoT Devices} have deployed highly scalable implementations that support processing requests and reading data based on different connected sensors.

\begin{figure}
    \centering
    \includegraphics[width=0.99\linewidth]{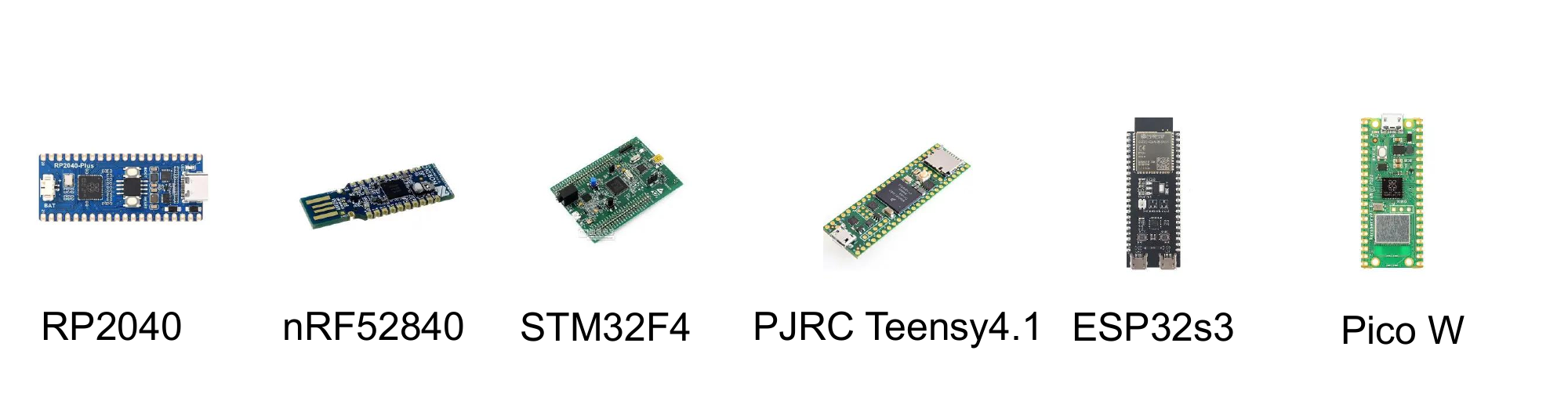}
    \caption{The 6 MCUs passed test in \name, each from different MCU family.}
    \label{mcus}
\end{figure}

This separation ensures both system efficiency and operational robustness, allowing for scalable deployment across a variety of IoT devices with different microcontroller units (MCUs) and sensor peripherals, and application scenarios.
Overall, MCP-IoT supports six MCU families and 22 sensors, which are listed in Figure~\ref{mcus} and Table~\ref{tab:Func} alongside their functionalities.

To provide an objective and comprehensive evaluation of the performance of {\name}, we also develop \textbf{{\name} Bench}, a benchmark suite that comprises {1,254} tasks and 3 system performance metrics, such as Tool Execution, Response Time, Memory Usage, and Prompt Robustness.
This benchmark systematically assesses various aspects of IoT task execution, including sensor data interpretation, device control accuracy, and system response reliability.
Through extensive experiments and using the developed {\name} Bench, {\name} achieves 100\% accuracy on Tool Execution Performance, 99\% accuracy on Prompt Robustness Assessment, with 205 ms average response time and 74 KB average memory usage, proving strong concurrency and scalability. 

\emph{Both the design of {\name} design and the {\name} Bench are open-sourced at~\url{https://github.com/Duke-CEI-Center/IoT-MCP-Servers}}.

\begin{table}[!t]
\centering

\caption{List of the 22 sensors used in {\name} with their functionalities. For sensors with more than one function, one special function, ``read-all'', is provided.}
\small
\begin{tabular}{|c|c|c|c|}
\hline
\textbf{Sensor} & \textbf{Functions} & \textbf{Sensor} & \textbf{Functions} \\
\hline
LTR390 & UV \& light & HW080 & detection \\
\hline
HC-SR501 & motion & TTP223 & touch \\
\hline
WaterSensor & water & GY302 & light \\
\hline
SW420 & vibration & HC-SR04 & distance \\
\hline
LM35 & temp. \& light & KY018& light \\
\hline
KY004 & button & KY020 & tilt \\
\hline
KY037 & touch \& light & KY023 & joystick \\
\hline
KY021& reed & KY038 & sound \\
\hline
KY010 & light & KY026 & flame \\
\hline
KY036 & touch & KY035 & field \& switch \\
\hline
DHT11 & \multicolumn{3}{|c|}{temperature \& humidity} \\
\hline
MPU6050 & \multicolumn{3}{|c|}{temperature \& angle \& acceleration \& gyro. } \\
\hline

\end{tabular}

\label{tab:Func}
\end{table}

\section{Related Work}

\myparatight{Tool calling and MCP.}
The Chain-of-Thought (CoT)~\cite{wei2023chainofthoughtpromptingelicitsreasoning} approach demonstrated that structured reasoning enhances the performance of LLMs, laying the foundation for subsequent LLM-based tool invocation. Follow-up works, such as Chameleon~\cite{lu2023chameleonplugandplaycompositionalreasoning} and Gorilla~\cite{patil2023gorillalargelanguagemodel}, further improved model capabilities by integrating external tools via API calls. Authors of~\cite{hsieh2023tooldocumentationenableszeroshot} subsequently proved that zero-shot prompting with tool documentation is sufficient to elicit accurate tool usage. Against this backdrop, MCP was proposed as a unified protocol for contextual tool documentation transmission.

\myparatight{MCP-based IoT integration.}
Recent work has explored the integration of MCP with IoT systems, particularly in the sensor domain. SensorMCP~\cite{Guo2025} presents a framework for automated sensor tool generation and language-driven sensor operation through MCP interfaces. Their approach enables large language models to automatically generate and operate sensor tools via a tool-language co-development pipeline, demonstrating effectiveness in scenarios such as wildlife monitoring systems. However, SensorMCP exhibits several limitations that constrain its practical applicability. First, the framework shows strong dependency on specific sensor types, limiting its extensibility to diverse IoT environments. Second, the integration of model training procedures within the framework results in prolonged response times, requiring approximately 30 minutes to accommodate new requirements. These limitations highlight the need for more scalable and responsive MCP-based IoT solutions.

\myparatight{MCP performance evaluation.}
Traditional LLM evaluation frameworks mainly focus on language comprehension abilities, but with the development of tool-enhanced AI, benchmarks specifically designed to evaluate tool usage abilities have become particularly important. Existing evaluation methods attempt to establish multi-dimensional evaluation standards~\cite{li2023apibankcomprehensivebenchmarktoolaugmented, xu2023toolmanipulationcapabilityopensource}, but there is no control over a unified context transfer protocol.

The evaluation of MCP-based systems has been addressed by MCP-RADAR~\cite{gao2025mcpradarmultidimensionalbenchmarkevaluating}, which introduces the first comprehensive benchmark for assessing LLM performance within the MCP framework. The benchmark employs a five-dimensional evaluation approach, providing valuable insights into LLM capabilities with standardized tool integration frameworks. While MCP-RADAR offers important evaluation metrics, it primarily focuses on assessing model capabilities rather than MCP system performance itself. This focus stems from the inherent challenge that different MCP servers execute distinct functions, making direct comparisons difficult. Moreover, system-level metrics such as operational efficiency and accuracy are challenging to evaluate directly within the existing framework.

\myparatight{Limitations of existing approaches.}
Current MCP-based IoT solutions face several critical limitations. First, existing frameworks lack comprehensive hardware compatibility, often targeting specific sensor types or device categories. Second, the integration of training procedures within operational frameworks introduces significant latency, hindering real-time IoT applications. Third, evaluation methodologies primarily assess model performance rather than system-level effectiveness, leaving gaps in understanding the MCP framework's performance in IoT contexts.

Our work addresses these limitations by proposing IoT-MCP, a decoupled architecture that separates connection management from MCP communication, and IoT Bench, a specialized benchmark that evaluates both model capabilities and system performance in the context of IoT systems.

\section{Architecture}

\subsection{Design Challenges}

The proposed {\name} framework aims to facilitate seamless integration between MCP and IoT systems, enabling LLMs to effectively interact with a wide range of IoT devices. 
However, the design of {\name} presents three fundamental challenges that must be addressed as illustrated below.

\myparatight{Response time on various MCU devices.}
{\name} acting as protocol adapters between LLMs and physical devices,  interface directly with language models, necessitating rapid response times to maintain conversational flow.  However, MCUs and sensors may experience temporary disconnections or hardware failures, potentially causing processing pipelines to stall. This mismatch between the high-availability expectations of LLM interactions and the inherent unreliability of distributed IoT devices creates a significant architectural challenge.

\myparatight{Heterogeneous sensory data management.}
The diversity of IoT sensors generates sensor-collected data in varied formats with distinct storage and processing requirements. These operations may be deployed locally on MCUs or in the cloud, making it difficult to establish unified data handling protocols. This heterogeneity complicates the development of standardized interfaces and consistent data processing workflows.

\myparatight{Computational resource constraints on IoT and edge devices.}
MCUs typically operate with limited computational resources, including restricted memory, processing power, and energy budgets. These constraints make it challenging to handle concurrent requests or perform complex data processing tasks. The need to support multiple sensor types and communication protocols further strains these limited resources, requiring careful optimization of both hardware utilization and software efficiency.

\subsection{IoT-MCP}


To address these challenges, we decouple the system into three distinct domains: (\emph{i}) Local Host, (\emph{ii}) Datapool and Connection Server, and (\emph{iii}) IoT Devices, as illustrated in Figure~\ref{fig:iotmcp-overview}.

\myparatight{Local Host.}
The Local Host hosts the LLMs and multiple specialized MCP servers within a controlled computing environment.
It maintains direct communication with the MCP clients while remaining isolated from potential disruptions in the IoT Devices. 
Each MCP server is designed to control a specific sensor function, allowing for more precise tool selection and execution with improved overall system reliability. Specifically, when the user enters a natural language request and calls the MCP Server, the following JSON command will be generated on the Local Host.
\begin{minted}[fontsize=\small]{json}
{
    "command": [READ_XXX],
    "duration": [DURATION],
    "interval": [INTERVAL]
}
\end{minted}
where \textsf{``duration''} refers to the time duration of one reading, and \textsf{``interval''} refers to the time interval between two readings.
This instruction will be passed to the IoT devices.

\myparatight{Datapool and Connection Server.}
This domain serves as an intermediary layer, which can be deployed either on the local host or in the cloud, depending on application-specific requirements such as:
\begin{itemize}
\item \textit{Deployment scale}: Local deployment suits small-scale networks (e.g., 10 MCUs)
\item \textit{Access demand}: Cloud deployment supports massive concurrent data requests
\item \textit{Resource constraints}: Offloads processing from resource-limited MCUs
\end{itemize} 
This flexibility addresses the heterogeneous sensory data management challenge by providing a standardized interface for data collection, storage, and analysis. The Datapool and Connection Server includes a \emph{connection server} that manages communication with MCU devices, and a \emph{datapool} for persistent data storage.
In Connection Server, each instruction will be assigned a unique ID for unified management. By buffering the requests on the Connection Server, this architecture avoids the temporary MCU disconnections from impacting the responsiveness of the Local Host.

\myparatight{IoT Devices.}
Each IoT device features a lightweight, extensible microservice architecture specifically designed for computational resource-constrained environments.
Each MCU runs a minimal service framework that supports multiple communication protocols that can either be wireless (e.g., Wi-Fi, Bluetooth) or wired (e.g., I2C~\cite{nxp:i2c}). After receiving the instruction, the MCU will generate the following JSON information and send it back to the Local Host.
\begin{minted}[fontsize=\small]{json}
{
    "write_time": [REAL_WORLD_TIME], 
    "timestamp": [TIMESTAMP],
    "id": [UUID], 
    "sensor": [SENSOR_NAME],
    [DATA_TYPE]: [DATA]
}
\end{minted}
Here, the field \textsf{[DATA\_TYPE]} is the returned sensory data, depending on the sensors (e.g., temperature).
The microservice design allows for dynamic extension based on connected sensors, enabling efficient resource utilization while maintaining broad hardware compatibility. This approach addresses computational constraints by distributing processing loads and allowing for modular functionality expansion.

\subsection{IoT-MCP Bench}

\subsubsection{Evaluation metrics}
The IoT-MCP Bench comprises two core components to validate the {\name} architecture.
First, a specialized dataset containing 114 \emph{Basic Tasks} representing fundamental sensor operations, and 1,140 procedurally-generated \emph{Complex Tasks} with linguistic variations.
Second, {\name} Bench, a dedicated testing toolkit. It is designed to systematically assess the framework's performance across
three key metrics: (\emph{i}) Success Rate, (\emph{ii}) Average Response Time, and (\emph{iii}) Peak Memory Footprint 
Based on these metrics, the evaluation framework focuses on seven dimensions directly aligned with experimental validation goals.
For example, tool execution reliability measures the task success rates and data accuracy across all supported sensors.
Response time analysis measures end-to-end latency while also isolating the time distribution across system components, including dedicated measurement of network connection overhead through idle response tests.
Memory utilization assessment tracks peak consumption during operations and idle states to characterize resource allocation patterns.

Cross-model compatibility testing evaluates performance variance across different language model backends using the tool execution success rate as the primary metric.
Concurrency performance analysis measures response times and memory usage under increasing simultaneous request loads. Prompt robustness assessment employs the Complex Tasks dataset to quantify success rates with linguistically challenging inputs.
Finally, deployment stability validation conducts extended real-world operation monitoring with automatic reconnection testing during simulated network interruptions.

\subsubsection{Dataset Generation}
\label{Dataset Generation}

The IoT Bench dataset is constructed through a two-stage hybrid generation approach that combines human expertise with automated complexity enhancement to create a comprehensive and challenging evaluation dataset.

\begin{figure}
    \centering
    \includegraphics[width=0.98\linewidth]{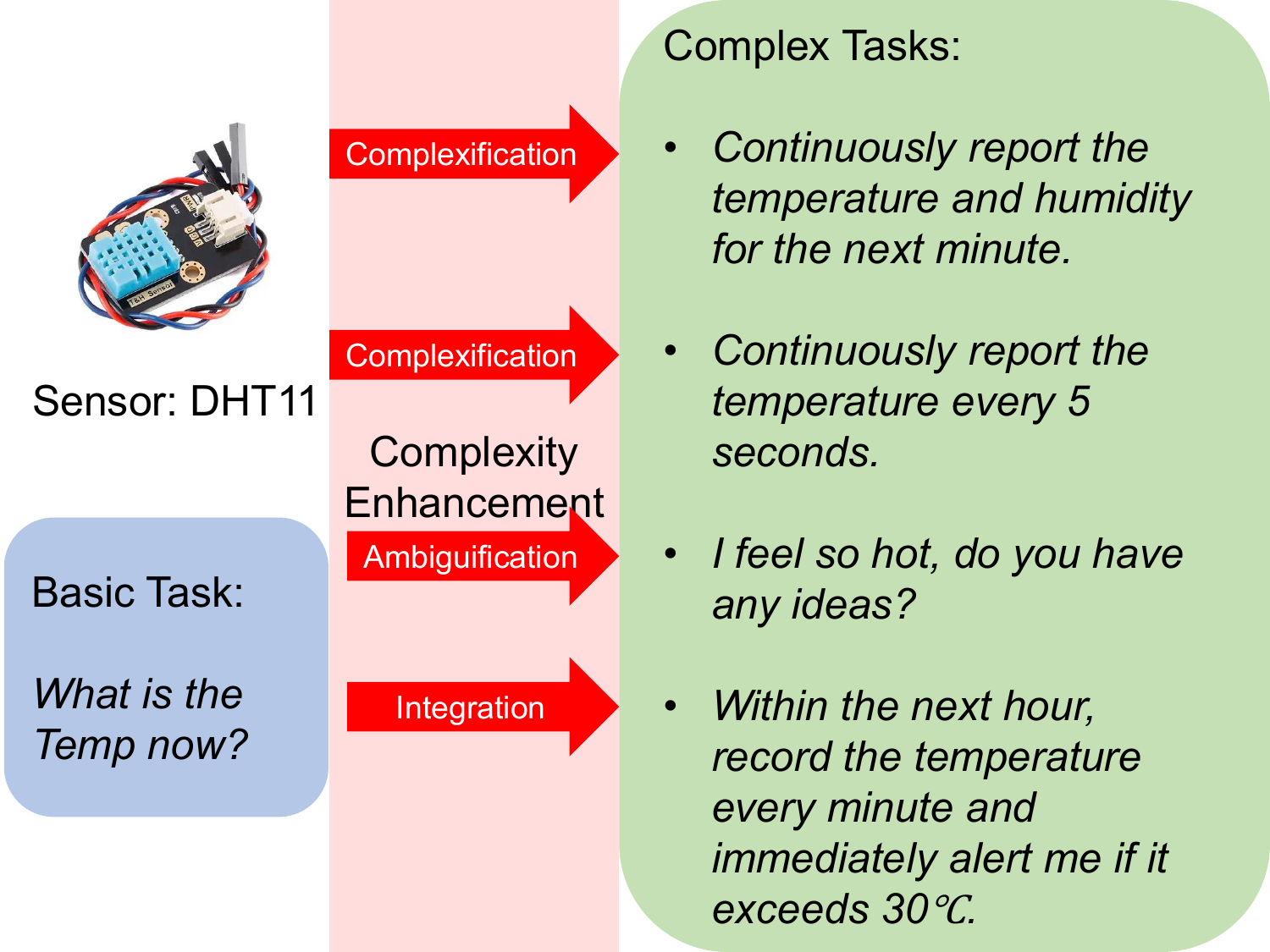}
    \caption{Example of complexity enhancement: Using the DHT11 sensor as an example, a simple temperature reading task is progressively transformed into a sequence of increasingly complex tasks.}
    \label{fig:tasks}
\end{figure}

\myparatight{Stage 1: Basic Task Generation.}
We begin by manually crafting 114 basic tasks based on the capabilities of 22 commonly used sensors supported by the {\name} framework.

These sensors support a variety of functionalities, including environmental monitoring (e.g., the DHT11 digital temperature and humidity sensor), motion detection (e.g., the MPU6050 6-axis accelerometer gyroscope sensor), imaging (e.g., camera), and specialized sensors (e.g., pH meter and gas monitor). 
Each fundamental task is designed to test specific sensor functionalities that will be invoked in different real-world IoT application scenarios.
The basic tasks are carefully designed to cover the full spectrum of sensor operations, including sensor initialization, configuration, and data reading.

\myparatight{Stage 2: Complexity Enhancement.}
To create a more challenging and realistic evaluation environment, we employ LLMs to systematically enhance and expand the 114 basic tasks through three transformation strategies.
This process generates 1,140 complex tasks (10 variants per basic task), resulting in a comprehensive dataset that spans simple single-sensor operations to complex multi-device orchestration scenarios. The dataset includes tasks with varying levels of ambiguity, different terminology choices, and diverse complexity levels, ensuring robust evaluation across different usage patterns and user expertise levels.

Specifically,
\textit{Complexification} involves combining multiple sensor operations into composite tasks that require sequential or parallel execution across different devices. 
\textit{Ambiguification} introduces natural language variations and implicit requirements that test the system's ability to interpret user intent from less structured inputs. 
\textit{Integration} creates scenarios that require coordination between multiple sensors and data fusion operations, reflecting real-world IoT application complexity. An example task is provided in Figure~\ref{fig:tasks}.

\section{Experiments and Results}


\begin{figure}
    \centering
    \includegraphics[width=1\linewidth]{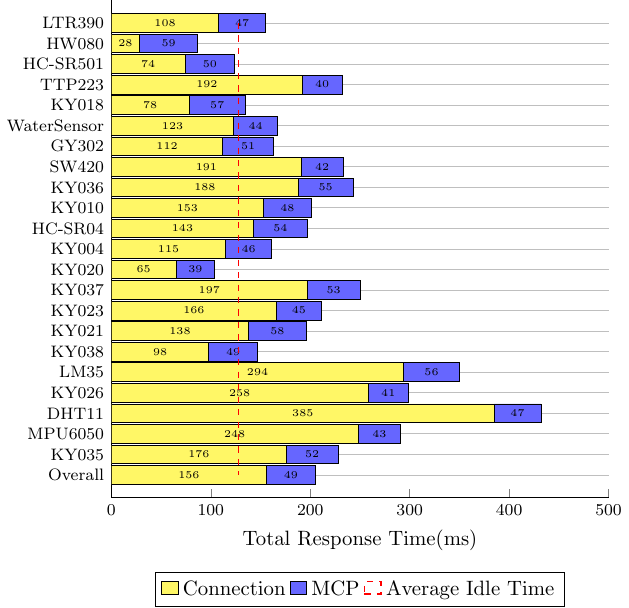}
    \caption{Measured average response time for the MCP servers. The yellow part represents the parsing and transmission time in the Connection Server and MCU, and the blue part represents the time consumption of the MCP Server. The red dashed line represents the average idle time.}
    \label{time}
\end{figure}
\begin{figure}
    \centering
    \includegraphics[width=1\linewidth]{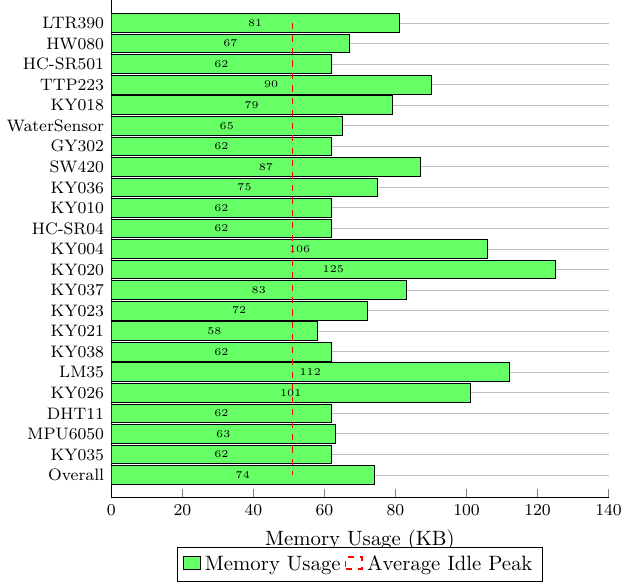}
    \caption{Peak memory footprint on MCU side. The red dashed line represents the average idle memory usage.}
    \label{memory}
\end{figure}

\subsection{Experimental Setup}

Unless otherwise specified, the following experiments are conducted using the Espressif ESP32-S3 microcontrollers~\cite{espressif_esp32_product} as the primary MCU platform, Claude~3.5 Haiku~\cite{Claude35Sonnet} as the base LLM, WiFi as the basic connection channel, and the Basic Task dataset described in Section ~\ref{Dataset Generation} 
All experiments are performed under controlled conditions to ensure reproducibility and statistical significance.
All MCP Servers' designs we have released and their respective supported features are shown in Table~\ref{tab:Func}.

\subsection{Metrics Evaluation}

\myparatight{Tool execution performance.}
To evaluate the reliability of the developed {\name} framework, we conduct comprehensive testing of the tool execution success rates and data accuracy across all supported sensors.
Specifically, each Basic Task is independently executed 10 times to establish statistical confidence, with results aggregated across different sensor categories and operation types.
The results show that all tasks can be successfully executed.

\myparatight{Response time analysis.}
System responsiveness is critical for maintaining interactive user experiences in IoT applications. We measure end-to-end response times and analyze time distribution across system components through repeated 10 times execution of the Basic Task dataset.

As illustrated in Figure~\ref{time}, the system achieves an average response time of 205 milliseconds. The analysis reveals that sensors utilizing I2C bus communication, particularly the MPU6050 accelerometer/gyroscope module, exhibit notably longer response times. Furthermore, our component-level timing analysis identifies the Connection Server as the primary performance bottleneck, accounting for approximately {30--75\%} of the total response time. This finding suggests that network communication and protocol translation represent the most significant latency sources in the current implementation.
Concurrently, in order to differentiate between time overhead attributed to network connection and core time overhead associated with sensor data reading, the idle response time test was repeated ten times, yielding an average value of 128 ms.

\begin{figure}
    \centering
    \includegraphics[width=0.98\linewidth]{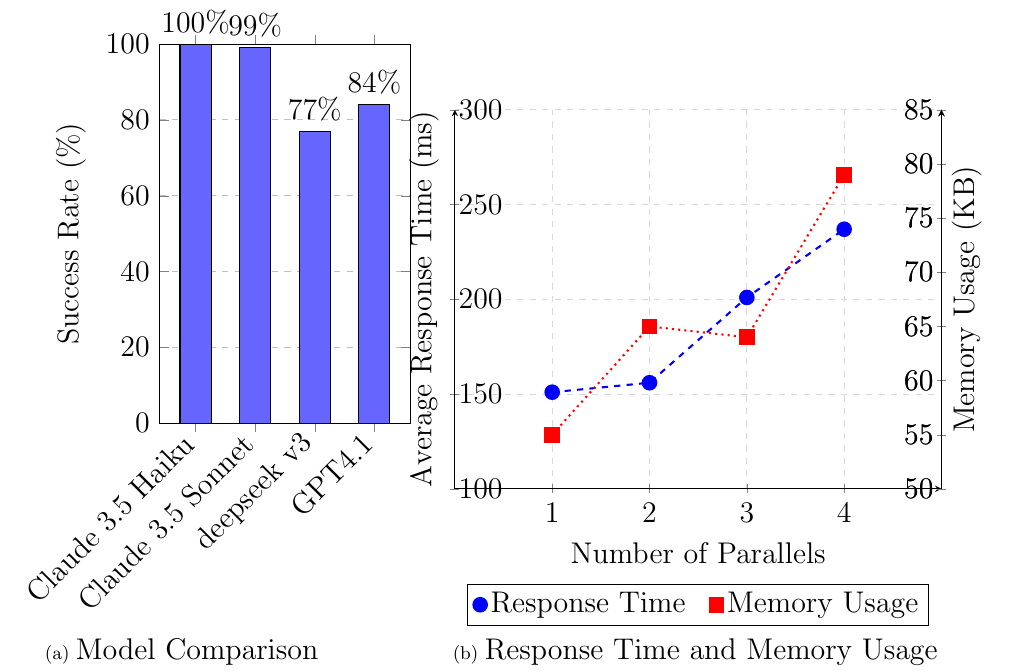}
    \caption{(\emph{Left}) Success rate of calling the MCP Server for different models; (\emph{Right}) Response time under multiple concurrent tasks.}
    \label{lr}
\end{figure}

\myparatight{Memory utilization characteristics.}
We monitor peak memory consumption across all system components during 10 times dataset execution, capturing both individual sensor requirements and system-wide memory usage patterns.
Figure~\ref{memory} demonstrates that the system maintains an average peak memory consumption of 74 KB across all tested scenarios. The results indicate relatively balanced memory allocation across different sensor implementations, with no single sensor type dominating system memory requirements. 
In a similar manner to the methodology employed for the idle time recording test, the idle memory usage test was repeated ten times, resulting in the attainment of an average value of 51 KB. This accounts for approximately {40--80\%} of the average peak usage. This finding indicates that the primary factor contributing to memory consumption in the MCU is the establishment of stable TCP connections. This observation suggests that the probability of our implementation encountering memory bottlenecks under high concurrency is low.

\begin{figure}
    \centering
    \includegraphics[width=0.98\linewidth]{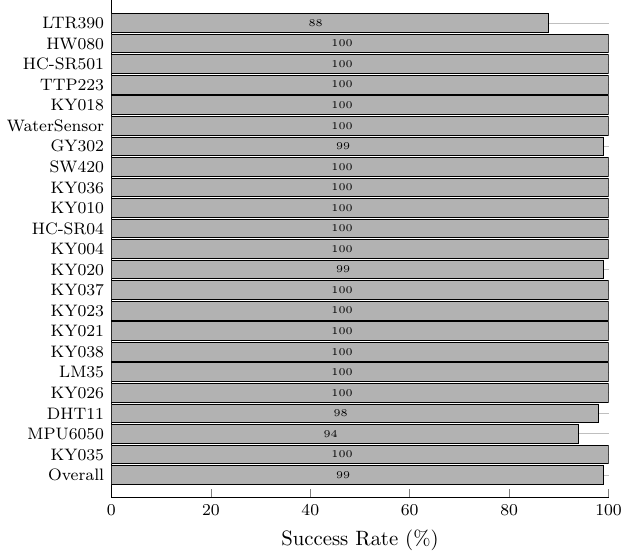}
    \caption{Success rate of the developed MCP Servers with complex tasks.}
    \vspace{-3mm}
    \label{complex}
\end{figure}

\subsection{System Robustness Evaluation}

\myparatight{Model robustness.}
To assess the generalizability of our framework across different language models, we evaluate system performance using alternative base models, including Claude 3.5 Sonnet~\cite{Claude35Sonnet}, DeepSeek V3~\cite{yangDeepSeekV3Advanced2024}, and GPT-4.1~\cite{GPT41}. This evaluation focuses on tool execution success rates as the primary metric for cross-model compatibility.

Results presented in Figure~\ref{lr}(\emph{Left}) demonstrate that our IoT-MCP Server implementations exhibit optimal compatibility with Claude models, achieving the highest success rates ({99--100\%}) in this configuration. Performance with alternative models shows modest degradation, with success rates decreasing by approximately 77\% for DeepSeek V3 and 84\% for GPT-4.1. This variation primarily stems from differences in tool calling conventions and parameter interpretation strategies across different model architectures.

\myparatight{Concurrency performance analysis.}
Real-world IoT deployments often require handling multiple simultaneous requests. To evaluate system behavior under concurrent load, we select the KY010 and KY036 sensors based on the experimental results presented earlier in this section as representative test cases and measure average response times and memory usage under varying concurrency levels.

Figure~\ref{lr}(\emph{Right}) illustrates the system's response to increasing concurrent request loads. The results demonstrate graceful performance degradation with increasing concurrency, maintaining acceptable response times ({150--250}\thinspace{ms}) even under high-load conditions. It was also observed that the growth brought about by high concurrency resulted in smoother memory usage ({55--79}\thinspace{KB}). The system exhibits linear scaling characteristics up to four concurrent requests.

\begin{figure*}[t]
    \centering
    \includegraphics[width=1\linewidth]{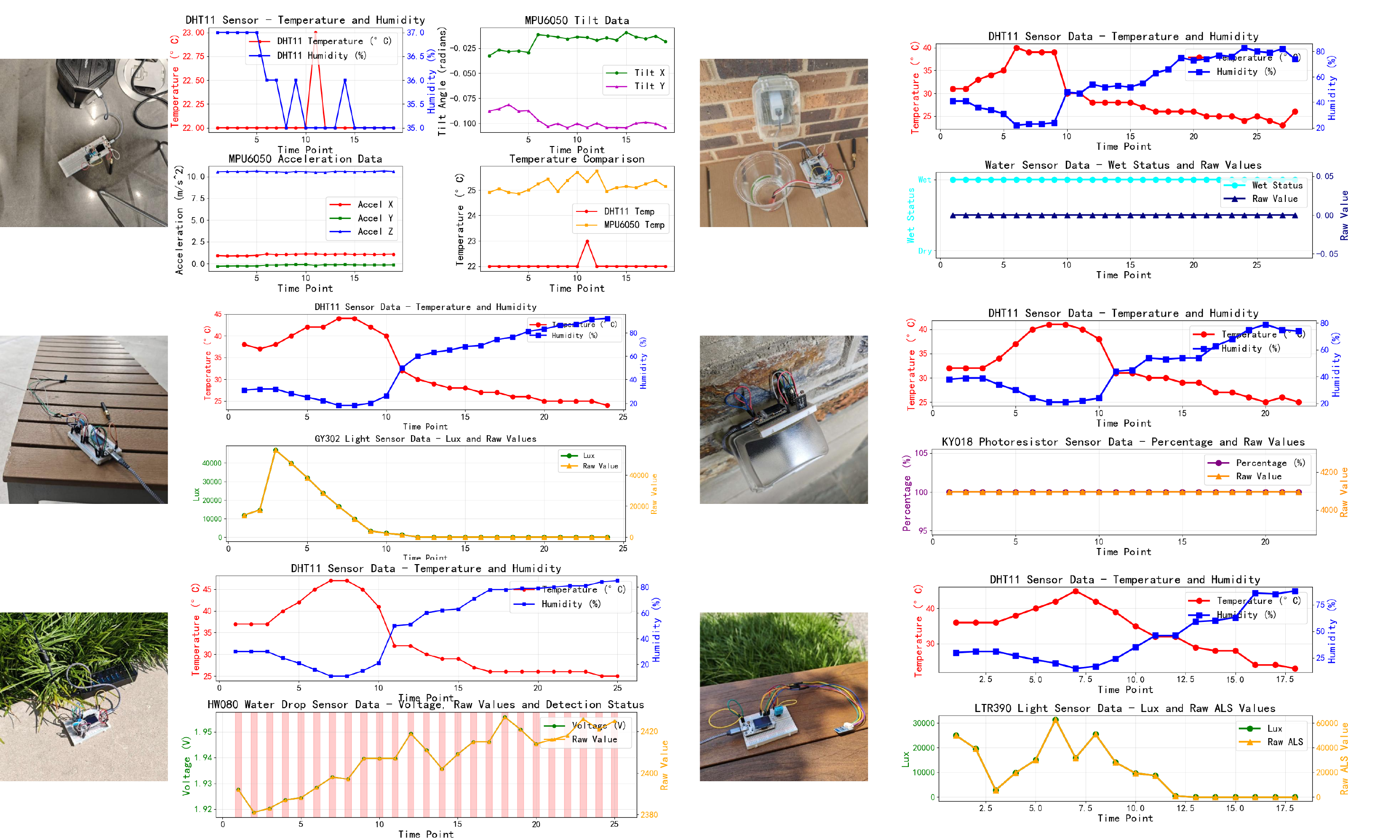}
    \caption{Data recorded in read-world deployment tests. 13 sensors (7 types) are connected to 6 MCUs and uniformly connected to {\name} for a 12-hour test.}
    \label{real exp}
\end{figure*}

\myparatight{Prompt robustness assessment.}
In order to evaluate system performance under actual usage conditions and respond to diverse and potentially ambiguous user inputs, a complex task dataset is generated through the fuzzification and complexification process. This complex task dataset is then subjected to repeated testing on each MCP Server until 100 tasks are completed. This assessment tests the framework's ability to interpret and execute commands despite linguistic variations.

As shown in Figure~\ref{complex}, the system maintains high performance even with challenging, ambiguous inputs, achieving an overall success rate of 99\%. It was observed that the servers that exhibited substandard performance (e.g., LTR390 and MPU6050) were precisely those that supported multiple data reads, while the tool instructions that generated errors were frequently designated as 'read-all', a practice that should have been avoided. This robust performance demonstrates the effectiveness of our natural language processing approach and validates the framework's readiness for deployment in diverse real-world scenarios where user inputs may not follow strict formatting conventions.

\subsection{Real-World Deployment Validation}

To validate the practical applicability and reliability of our framework, we conduct a comprehensive 12-hour deployment test within a multi-story building environment. The deployment consists of 6 ESP32-S3 microcontrollers equipped with 7 different types, 12 sensors, all connected to a WiFi network to simulate realistic IoT infrastructure conditions.

Figure~\ref{real exp} presents continuous monitoring results over the 12-hour period. As expected, all sensors returned continuous results. Based on the Connection Server's design, the IoT-MCP can maintain stable connections and automatically restore them after unexpected disconnection, such as during a power outage or network abnormality. This real-world validation confirms the framework's readiness for production deployment and validates the architectural decisions underlying the IoT-MCP design.

\section{Discussions}

\subsection{Hardware Scope Limitations}

The current framework's exclusive focus on sensor-based devices constrains applications to monitoring scenarios, omitting actuator integration. However, the decoupled architecture inherently supports extending to control mechanisms. Such expansion would enable closed-loop environmental control systems, transforming passive monitoring platforms into active intervention systems capable of executing complex sensor-actuator sequences for user feedback.

\subsection{Functional Complexity Limitations}
While supporting basic device invocation, the existing architecture lacks capabilities for dynamic workflow composition, a limitation rooted in positioning LLM+MCP clients as mere \textit{callers} rather than \textit{designers}. Future work will reposition clients as system designers through four interconnected capabilities: composition engines for generating execution plans; workflow management systems with robust failure handling; performance-based optimization mechanisms; safety protocols ensuring graceful degradation during operation.

\section{Conclusions}
We presented {\name}, a novel framework bridging MCP and IoT systems to address scalable control challenges.
Specifically, the design of {\name} is decoupled into three domains: (\emph{i}) Local Host, (\emph{ii}) Datapool and Connection Server, and (\emph{iii}) IoT Devices.
We also introduced a comprehensive benchmark dataset comprising 1,254 tasks that span 22 sensors and 6 MCUs from different families, systematically evaluating {\name} across (\emph{i}) Success Rate, (\emph{ii}) Average Response Time, and (\emph{iii}) Peak Memory Footprint. In summary, {\name} represents a significant advancement toward real-world LLM-IoT integration in the era of LLMs.

\section*{Acknowledgments}
The work was supported in part by NSF grants CNS-2112562 and CNS-2330333.

\bibliographystyle{ACM-Reference-Format}
\bibliography{ref}
\end{document}

%% file: macro.tex
\definecolor{lightgray}{gray}{0.9}
\definecolor{lightblue}{rgb}{0.9,0.9,1}
\definecolor{LightMagenta}{rgb}{1,0.5,1}
\definecolor{red}{rgb}{1,0,0}

\newcommand{\remove}[1]{}

\newcommand{\myparatight}[1]{\vspace{0.5ex}\noindent\textbf{#1~~}}

\newcommand\name{{IoT-MCP}}
\newcommand\namebf{{\textbf{IoT-MCP}}}